\documentclass[reprint,amsmath,amssymb,aps]{revtex4-1}

\usepackage[T1]{fontenc}
\usepackage{nicefrac}
\usepackage{color}
\usepackage{hyperref}

\newcommand{\psec}[1]{\emph{#1.---}}

\newcommand{\emm}{\mathcal{M}}
\newcommand{\uui}{\mathcal{U}_i}
\newcommand{\Mpl}{M_{\rm Pl}}

\begin{document}

\title{On the cosmological constant problem}

\author{Lucas Lombriser}

\affiliation{D\'{e}partement de Physique Th\'{e}orique, Universit\'{e} de Gen\`{e}ve, \\ 24 quai Ernest Ansermet, 1211 Gen\`{e}ve 4, Switzerland}

\date{\today}

\begin{abstract}

An additional variation of the Einstein-Hilbert action with respect to the Planck mass provides a constraint on the average Ricci scalar that prevents vacuum energy from gravitating.
Consideration of the evolution of the inhomogeneous matter distribution in the Universe with evaluation of the averaging constraint on disconnected matter cells that ultimately form isolated gravitationally bound structures yields a backreaction effect that self-consistently produces the cosmological constant of the background.
A uniform prior on our location in the formation of these isolated structures implies a mean expectation for the present cosmological constant energy density parameter of $\Omega_{\Lambda}=0.704$, giving rise to a late-time acceleration of the cosmic expansion and a coincident current energy density of matter.

\end{abstract}

\maketitle

\psec{Introduction}
%
The physical nature of the cosmological constant remains a persistent enigma immanent to Einstein's Theory of General Relativity.
It is generally thought to represent the gravitational contribution of vacuum fluctuations, anticipated of adequate magnitude to account for the observed late-time accelerated expansion of our Universe~\cite{Riess:1998cb,Perlmutter:1998np}.
Quantum theoretical calculations, however, exceed measurement by several orders of magnitude~\cite{Weinberg:1988cp,Martin:2012bt}.
This may imply a missing prescription for the correct computation of standard vacuum contributions but also motivates conjectures of an undetermined mechanism that prevents vacuum energy from gravitating in full extent and the possibility of attributing cosmic acceleration to a different origin.
The growing wealth of cosmological observations, however, also puts strong constraints on alternatives to a cosmological constant as explanation of the accelerated expansion such as dark energy or a departure from General Relativity at large scales~\cite{Aghanim:2018eyx,Abbott:2017wau,Lombriser:2016yzn,Monitor:2017mdv}.
A curiosity of cosmic acceleration can furthermore be found in the comparable magnitude of its associated current energy density with that of matter, provoking the \emph{Why Now?} conundrum~\cite{Martin:2012bt,Lombriser:2017cjy}.

This \emph{Letter} re-examines the cosmological constant problem under an additional variation of the gravitational and matter actions with respect to the Planck mass, which allows for an interpretation of the Planck mass as global Lagrange multiplier that imposes general relativistic dynamics on the metric prescribing the space-time for the matter fields.
The resulting additional constraint equation is evaluated respecting the inhomogeneous nature of the Universe at small scales, and an analysis is presented for the implication of this new framework for the \emph{old} cosmological constant problem of the non-gravitating vacuum as well as its \emph{new} aspects of cosmic acceleration and the coincidence problem.

\psec{Non-gravitating vacuum energy}
%
Consider the Ein-stein-Hilbert action
\begin{equation}
 S = \frac{\Mpl^2}{2} \int_{\emm} d^4x \sqrt{-g} \left( R - 2\Lambda \right) + \int_{\emm} d^4x \sqrt{-g} \mathcal{L}_{\rm m} + b.t. \,, \label{eq:action}
\end{equation}
where $\emm$ denotes the cosmic manifold, $\Lambda$ is a free classical cosmological constant, matter fields are minimally coupled, the standard Gibbons-Hawking-York boundary term is adopted, and
$c=\hbar=1$.
Variation of the action with respect to the metric $g_{\mu\nu}$ yields the Einstein field equations
\begin{equation}
 G_{\mu\nu} + \Lambda g_{\mu\nu} = \Mpl^{-2} T_{\mu\nu} \,, \label{eq:einstein}
\end{equation}
where $T_{\mu\nu}\equiv-2\left[ \delta(\sqrt{-g}\mathcal{L}_{\rm m})/\delta g^{\mu\nu} \right]/\sqrt{-g}$.
In addition to the metric variation, we shall now perform a variation of the action with respect to the quadratic Planck mass $\Mpl^2$.
This may be interpreted as using $\Mpl^2$ as a global Lagrange multiplier for a topological constraint on the the matter action. Boundary conditions may be adapted as in Ref.~\cite{Kaloper:2016yfa} (also see Ref.~\cite{BeltranJimenez:2017tkd}).
This gives the constraint equation
\begin{equation}
 \frac{1}{2}\int_{\emm} d^4x \sqrt{-g} (R - 2\Lambda) = 0 \,. \label{eq:constraint}
\end{equation}
Using the trace of Eq.~\eqref{eq:einstein} in Eq.~\eqref{eq:constraint}, it follows that
\begin{equation}
 \Mpl^2\Lambda = \frac{1}{2} \frac{\int_{\emm} d^4x \sqrt{-g} T}{\int_{\emm} d^4x \sqrt{-g}} \equiv \frac{1}{2} \langle T \rangle \label{eq:lambda}
\end{equation}
such that the Einstein field equations~\eqref{eq:einstein} become
\begin{equation}
 G_{\mu\nu} + \frac{1}{2}\Mpl^{-2}\langle T \rangle g_{\mu\nu} = \Mpl^{-2} T_{\mu\nu} \,, \label{eq:einstein2}
\end{equation}
replacing the free cosmological constant with a space-time average of the trace of the energy-momentum tensor.

This new metric field equation is reminiscent of vacuum energy sequestering~\cite{Kaloper:2013zca,Kaloper:2015jra} but with a different fraction of $\Mpl^{-2}\langle T \rangle$ contributing dynamically as the cosmological constant.
Similarly to the sequestering mechanism, vacuum contributions to the matter sector cancel out in Eq.~\eqref{eq:einstein}.
However, the cancellation occurs differently.
For this observation, consider first the one-loop vacuum cosmological constant in curved space-time~\cite{Martin:2012bt}
\begin{equation}
 \Lambda_{\rm vac} = \Mpl^{-2} \sum_i n_i \frac{m_i^4}{64\pi^2} \ln\left( \frac{m_i^2}{\mu_i^2} \right) + \Lambda_{\rm vac}^{\rm EW} + \ldots \,, \label{eq:oneloopvacuum}
\end{equation}
where $m_i$ denote particle masses of species $i$, $n_i$ represent the respective number of degrees of freedom with $+/-$ sign for bosons/fermions, and $\mu_i$ are unknown renormalization mass scales.
The electroweak vacuum contribution is given by $\Lambda_{\rm vac}^{\rm EW} = -\Mpl^{-2}(\sqrt{2}/16)(m_{\rm H}^2/G_{\rm F})$ with Higgs boson mass $m_{\rm H}$ and Fermi constant $G_{\rm F}$, and one may also wish to include further phase transitions.
As an example for cancellation, one may now rewrite the masses as fractions of the Planck mass $m_i=\lambda_i\Mpl$ and using the scale $\mu_i=m_i\exp(-M^2/\Mpl^2)$ for some renormalization mass $M$ independent of $\Mpl$, one finds that $\Mpl^2 \Lambda_{\rm vac} \propto \Mpl^2 M^2$.
The same scaling is, for instance, also found for the leading-order vacuum contribution of a Wheeler space-time foam description~\cite{Wang:2017oiy}.
Separating out the vacuum and bare component from the matter Lagrangian density, $\mathcal{L}_{\rm m} = \bar{\mathcal{L}}_{\rm m} - \Mpl^2 (\Lambda_{\rm vac} + \Lambda_{\rm bare})$, one finds after variation of the action with respect to $g_{\mu\nu}$ and $\Mpl^2$ that $G_{\mu\nu} + (\Lambda + \Lambda_{\rm vac} + \Lambda_{\rm bare})g_{\mu\nu} = \Mpl^{-2}\tau_{\mu\nu}$ and $\Mpl^2(\Lambda + \Lambda_{\rm vac} + \Lambda_{\rm bare}) = \langle \tau \rangle /2$, where $\tau$ is specified by $\bar{\mathcal{L}}_{\rm m}$,
and therefore,
\begin{equation}
 G_{\mu\nu} + \frac{1}{2}\Mpl^{-2}\langle \tau \rangle g_{\mu\nu} = \Mpl^{-2} \tau_{\mu\nu} \,. \label{eq:einstein3}
\end{equation}
Hence, the vacuum and bare cosmological constants do not gravitate.
Rather than a cancellation between the left- and right-hand sides of Eq.~\eqref{eq:einstein2}, as in vacuum energy sequestering, for given $\Lambda_{\rm vac}$ and $\Lambda_{\rm bare}$, the value of $\Lambda$ is set by the topological constraint equation~\eqref{eq:constraint} such that the sum of the cosmological constants matches $\langle \tau \rangle/2$.

Importantly, however, there is no guarantee that the residual cosmological constant obtained by the scaling of masses in the one-loop vacuum term~\eqref{eq:oneloopvacuum} is radiatively stable or that the vacuum contribution should even scale as $\Mpl^2$.
To simultaneously meet both caveats we shall first consider a scaling of the total vacuum energy density with the Planck mass as $\Mpl^{2\alpha}$ such that the total vacuum contribution to the action~\eqref{eq:action} will be $-\Mpl^{2\alpha} \bar{\Lambda}_{\rm vac}$ with the overbar indicating an independence of $\Mpl^2$.
In particular, this allows for $\alpha=0$.
What is necessary for the cancellation is the addition of a classical counterterm $-\Mpl^{2\alpha}\bar{\Lambda}_{\alpha}$.
More specifically, following the same procedure as for Eq.~\eqref{eq:einstein3} and solving for $\bar{\Lambda}_{\alpha}$, one obtains
\begin{equation}
 G_{\mu\nu} + \frac{1}{2-\alpha} \left[ (1-\alpha) \Lambda + \frac{\Mpl^{-2} \langle \tau \rangle}{2} \right]g_{\mu\nu} = \Mpl^{-2} \tau_{\mu\nu} \,, \label{eq:einstein4}
\end{equation}
where $\Lambda$ remains a free classical cosmological constant that is radiatively stable and determined by measurement.
As observed next from the evolution of matter inhomegeneities, $\Lambda$ simply corresponds to the total measured cosmological constant.
For $\alpha=1$, one recovers Eq.~\eqref{eq:einstein3}, and for $\alpha=0$, the dynamical equations of the local sequestering mechanism~\cite{Kaloper:2015jra} with $\Lambda_{\rm tot} = \nicefrac{1}{4} \langle \tau \rangle \Mpl^{-2} + \Delta\Lambda$, where $\Delta\Lambda\equiv\Lambda/2$.
Note that one may also consider a series expansion of $\Lambda_{\rm vac}$ in $\Mpl^2$, for instance from considering graviton loops~\cite{Kaloper:2016jsd}.
With a counterpart expansion of the classical cosmological constant, the resulting equivalent expression for Eq.~\eqref{eq:constraint} cancels these terms and reproduces Eq.~\eqref{eq:einstein4}. 
Similarly, quantum corrections with higher-derivative terms in Eq.~\eqref{eq:action} do not contribute to Eq.~\eqref{eq:constraint} or the field equations if independent of Planck mass, and if dependent on $\Mpl$, are cancelled by the same classical expansion (also see Ref.~\cite{Kaloper:2016jsd}).

Finally, variations with respect to the Planck mass have also been performed in Ref.~\cite{Kaloper:2015jra,Kaloper:2016jsd} and are in nature similar to a scalar-tensor theory in Jordan-Brans-Dicke representation with constant scalar field across the observable universe.
A transformation into Einstein frame removes the variation in $\Mpl^2$ but leaves variations with respect to an effective $\Lambda$ and a coupling in the matter sector, sharing similarities with the proposals of Refs.~\cite{Henneaux:1984ji,Unruh:1988in,Barrow:2010xt,Kaloper:2013zca}, however also differing from them, e.g., by not imposing a constant four-volume as in unimodular gravity.
The scalar field becomes a space-time constant, for example, by a $\delta$-function generated through appropriate boundary conditions on an additional vector field~\cite{Henneaux:1984ji,Shaw:2010pq} or by a squared four-form field strength contribution of a three-form gauge field as arises in supergravity~\cite{Aurilia:1980xj,Hawking:1984hk,Bousso:2000xa,Shaw:2010pq,Kaloper:2015jra,Kaloper:2016jsd}.
Variations in $\Mpl^2$ therefore find fundamental motivation ranging from scalar-vector-tensor or higher-dimensional scalar-tensor theories to supergravity, string theory, or a type II multiverse.

\psec{Backreaction from structure formation}
%
The four-volume term in Eq.~\eqref{eq:lambda} determining the residual cosmological constant in Eq.~\eqref{eq:einstein2} grows large for an old universe, and with integration over the background matter density $\bar{\rho}_{\rm m}$, assumed spatially perfectly homogeneous and isotropic, $\langle \tau \rangle = \langle \bar{\rho}_{\rm m} \rangle$ eventually vanishes.
For the residual to reproduce the observed cosmological constant with Planck parameters~\cite{Aghanim:2018eyx}, the Universe should have undergone an immediate collapse at the scale factor $a=0.926$, at an age of $0.88H_0^{-1}$, thus, about 1~Gyr in the past, and in contrast, an immediate collapse at the current epoch would account for 81\% of the observed value with a decreasing fraction for a longer future (cf.~\cite{Lombriser:2018aru}).
While it is interesting that this value is close to measurement, it is not an exact recovery and moreover standard cosmology does not foresee an imminent collapse of the Cosmos.

Importantly, however, the Universe is not perfectly homogeneous and isotropic with structures growing to be even more pronounced in the future.
We shall therefore next examine the impact of the evolution of inhomogeneities in the generation of the residual cosmological constant through Eq.~\eqref{eq:lambda}, thus a $\emph{backreaction}$ effect of structure formation.
To describe the inhomogeneous matter distribution in the Universe, consider a separation of the matter content into disconnected cells $\uui$, which are of maximal extent such that the matter contained will remain gravitationally bound throughout their evolution, eventually forming isolated clusters in the far future.
By a halo model interpretation, all matter is contained in these cells with each particle uniquely assigned to a single $\uui$.
The matter action in Eq.~\eqref{eq:action} becomes
\begin{equation}
 S_{\rm m} = \sum_i \int_{\uui} dV_4 \mathcal{L}_{{\rm m},i} + \int_{\emm\backslash\bigcup_i\uui} dV_4 \mathcal{L}_{\varnothing} \,,
\end{equation}
where $\mathcal{L}_{{\rm m},i}$ denote the Lagrangian densities of the matter in $\uui$ and $\mathcal{L}_{\varnothing}$ represents a Lagrangian density for the empty space.
$\mathcal{L}_{\varnothing}$ is chosen such that the residual cosmological constant in the empty space matches that of the matter cells.
We shall adopt $\mathcal{L}_{\varnothing} = \Mpl^2 \left( \Lambda/n + \Mpl^{2n} \bar{\Lambda}_{\varnothing} \right)$.
Variation of the new action with respect to the metric and $\Mpl^2$ on each $\uui$ gives
\begin{equation}
 \Lambda = \frac{1}{2\Mpl^2\rvert_{\uui}} \langle T \rangle_{\uui} \,, \label{eq:cellconstraints}
\end{equation}
where $\langle T \rangle_{\uui}$ indicate averages over the manifolds $\uui$.
Variations on $\emm\backslash\bigcup_i\uui$ do not impose a constraint on $\Lambda$ such that there is no conflict with the constraint obtained from the matter patches in Eq.~\eqref{eq:cellconstraints}.
It is easy to verify that for $n\rightarrow(\alpha-1)$ the vacuum and bare cosmological constants remain non-gravitating in any of the patches,
and as observed in the following by examining the evolution of critical mass shells existing at turnaround between expansion and collapse, $\langle \tau \rangle_{\uui} = \langle \tau \rangle_{\mathcal{U}_j} = 2\Mpl^2 \Lambda$ $\forall i, j$.

For simplicity, consider first the evolution and formation of nonlinear structures employing the spherical collapse model with approximation of forming halos by spherically symmetric top-hat overdensities and a further restriction to only pressureless matter and a cosmological constant in a spatially-flat cosmological background.
It has been checked in Ref.~\cite{Lombriser:2018aru} that radiation components only marginally affect the computation.
The top hat is defined by its density $\rho_{\rm m} \equiv \bar{\rho}_{\rm m} + \delta\rho_{\rm m} \equiv \bar{\rho}_{\rm m} (1 + \delta)$ and mass $M$.
The background density corresponds to the spatial cell volume integration of all top-hat densities with respect to the total spatial volume of the Universe, which follows from mass conservation and reproduces the standard Friedmann equations.
From energy-momentum conservation in each matter cell $\nabla^{\mu}T_{\mu\nu}\rvert_{\uui}=0$, one derives the evolution equation~\cite{Lombriser:2018aru}
\begin{equation}
 y'' + \left( 2 + \frac{H'}{H} \right) y' + \frac{1}{2} \Omega_{\rm m}(a) \left( y^{-3} - 1 \right) y = 0 \label{eq:y}
\end{equation}
for the dimensionless physical top-hat radius $y=(\rho_{\rm m}/\bar{\rho}_{\rm m})^{-1/3}$, where primes denote derivatives with respect to $\ln a$ and $\Omega_{\rm m}(a) \equiv \Mpl^{-2} \bar{\rho}_{\rm m}/(3H^2)$ with Hubble function $H$.
The evolution of $y$ can be determined setting initial conditions in the matter-dominated regime $a_i\ll1$, where $y_i \equiv y(a_i) = 1 - \delta_i/3$ and $y_i' = - \delta_i/3$ for an initial top-hat overdensity $\delta_i$.
Note that for $y\approx1$, Eq.~\eqref{eq:y} reduces to the familiar differential equation determining the growth of linear matter density perturbations in $\Lambda$CDM.

Evaluating Eq.~\eqref{eq:lambda} on a given patch $\uui$, one finds~\cite{Lombriser:2018aru}
\begin{equation}
 \langle \tau \rangle = \frac{\int d^4x \sqrt{-g} \: \tau}{\int d^4x \sqrt{-g}} = \bar{\rho}_{\rm m0} \frac{\int d\ln a \: H^{-1}}{\int d\ln a \: H^{-1} a^3 y^3} \,, \label{eq:tauevaluate}
\end{equation}
where subscripts of zero denote present values.
Importantly, the evolution of $y$, and therefore the value of $\langle \tau \rangle$, depends only on the initial overdensity $\delta_i$ and is independent of the top-hat mass.
From the competition between the expansion of the cosmological background and the self gravity of the massive cell one can therefore find a universal minimal, critical $\delta_i$ below which the expansion rate in the future will exceed the effect of self gravitation and above which a given patch will collapse.
Hence, for critical mass shells $\langle \tau \rangle = \langle \tau \rangle_{\uui} = \langle \tau \rangle_{\mathcal{U}_j}$ $\forall i, j$.
Using the symmetry of $a(t)y(t)$ around $t_{\rm turn}=t_{\rm max}/2$ for mass shells that eventually collapse, one furthermore obtains
\begin{equation}
 \frac{\langle \tau \rangle}{\Mpl^2\Lambda_{\rm obs}} = \frac{\Omega_{\rm m}}{2(1-\Omega_{\rm m})} \frac{t_{\rm max}}{\int_0^{t_{\rm turn}} dt \; a^3 y^3} \,,
\end{equation}
where $\Lambda_{\rm obs}$ denotes the observed cosmological constant that drives the late-time acceleration of the background.
From Eq.~(\ref{eq:y}) it follows that $d (a y)/dt=0$ and $d^2 (a y)/dt^2=0$ at $t_{\rm turn}$ such that $a^3y^3|_{t_{\rm turn}}=\Omega_{\rm m}/(1-\Omega_{\rm m})/2$, and thus,
\begin{equation}
 \int_0^{t_{\rm turn}} dt \; a^3 y^3 < \frac{\Omega_{\rm m}}{2(1-\Omega_{\rm m})} t_{\rm turn} \,, \label{eq:ayturn}
\end{equation}
implying that
\begin{equation}
 \frac{\langle \tau \rangle}{\Mpl^2\Lambda_{\rm obs}} > \frac{t_{\rm max}}{ t_{\rm turn}} = 2 \,. \label{eq:factortwo}
\end{equation}
The longer the evolution remains at $a^3y^3|_{t_{\rm turn}}$, which is the case for nearly critical matter patches that only collapse in the far future, the closer the ratio approximates this limit.
Hence, in Eqs.~\eqref{eq:einstein3} and \eqref{eq:einstein4} it follows that $\Lambda_{\rm obs} = \Mpl^{-2}\langle \tau \rangle/2 = \Lambda$ such that we recover the Einstein field equation $G_{\mu\nu} + \Lambda_{\rm obs} g_{\mu\nu} = \Mpl^{-2} \tau_{\mu\nu}$.
This result is independent of assuming a top-hat description for the disconnected matter cells.
Critical matter patches that exist long enough into the cosmological constant dominated regime always reach the constant value of $a\:y$ in Eq.~\eqref{eq:ayturn}~\cite{Dunner:2006rf} 
and hence reproduce Eq.~\eqref{eq:factortwo}.
For example, using the analytic description for the evolution of critical mass shells in Ref.~\cite{Dunner:2006rf} for the integration in Eq.~\eqref{eq:tauevaluate} recovers the limit in Eq.~\eqref{eq:factortwo}. 
In the very far future the critical cells will eventually collapse or fragment into smaller structures due to energy loss from radiation.
These isolated halos may then further collapse into ultra massive black holes that eventually evaporate.
It is also possible that the entire Universe will collapse due to a Higgs instability prior to that.
These processes are, however, expected to occur on much longer time scales than required for approaching the limit in Eq.~\eqref{eq:factortwo}.
But as a consequence, the integrals in Eq.~\eqref{eq:lambda} are not expected to diverge.
Importantly, even for an eternal integration, the limit in Eq.~\eqref{eq:lambda} remains well defined under l'H\^opital's rule with constant $\tau$ in the de Sitter future (also see Ref.~\cite{Kaloper:2015jra}).
It is also worth emphasizing that because of the integration over the entire existence in time, in contrast to similar approaches with causal set theory~\cite{Zwane:2017xbg} or backreaction from long-wavelength perturbations~\cite{Brandenberger:2018fdd}, $\langle \tau \rangle$ and $\Lambda_{\rm obs}$ in Eq.~\eqref{eq:factortwo} are time independent.

Finally, note that the recovery of the observed cosmological constant in the Einstein equations is not a circular argument.
A different fraction of $\Mpl^{-2}\langle \tau \rangle$ contributing as the residual cosmological constant in Eq.~\eqref{eq:einstein3} would not provide a self-consistent solution (cf.~\cite{Lombriser:2018aru}).
Moreover, as we will see next, the evaluation of $\langle \tau \rangle$ by the evolution of isolated critical matter patches offers direct implications for the coincidence problem.

\psec{Late-time acceleration and coincidence problem}
%
The additional variation of the Einstein-Hilbert action with respect to the Planck mass combined with the evolution of disconnected critical matter cells provides a self-consistent framework to alleviating the \emph{old} cosmological constant problem in preventing vacuum energy from gravitating.
The observational evidence for a late-time accelerated expansion~\cite{Riess:1998cb,Perlmutter:1998np} adds a \emph{new} aspect to the problem by the small non-vanishing $\Lambda_{\rm obs}$ that furthermore coincides with the current energy density of matter.
This \emph{Why Now?}~coincidence problem is not self-evidently addressed by a mechanism that gives rise to a free classical, radiatively stable cosmological constant.
While a late-time epoch of cosmic acceleration will generally be assumed in the presence of a non-vanishing positive cosmological constant, for an estimation of the relative magnitude of its current contribution to the total energy density, one needs a sense of likelihood of our particular location in the cosmic history.
One may, for instance, inspect the star-formation history, and allowing the Sun to have formed within $3\sigma$ of the star-formation peak constrains the current energy densities of matter and the cosmological constant to differ not beyond a factor of $10^2$~\cite{Lombriser:2018aru}.

The relation of the residual cosmological constant to isolated matter cells however provides a more direct estimate of $\Omega_{\Lambda}\equiv\Lambda/(3H_0^2)$ that is intrinsically linked to the evolution in Eq.~\eqref{eq:y}.
The dimensionless physical top-hat radius $y$ lends itself here to a meaningful measure of likelihood~\cite{Lombriser:2018aru}, describing the relative extent of the matter cells and evolving within a finite range from a small perturbation $-\delta_i/3$ from unity at early times to zero at the time of its collapse in the far future.
Adopting a uniform prior on $y$ yields a mean expectation value of $\langle y \rangle = \nicefrac{1}{2}$, which shall be assumed here as a likely location for our present time $t_0$, implying that there is as much of the evolution of $y$ ahead of us as there is in the past.
To compute $\Omega_{\Lambda}$ at $t_0$ one can adopt a more convenient normalization for the scale factor $a$ that eliminates the dependence of Eq.~\eqref{eq:y} on $\Omega_{\rm m}$.
As long as the cosmological constant is not strictly vanishing, there is always a time when $\bar{\rho}_{\rm m}=\bar{\rho}_{\Lambda}=\Mpl^2\Lambda$.
Hence, normalizing  the scale factor at the time of this equality, one therefore finds $H_{\rm eq} \equiv H(a=a_{\rm eq}\equiv1)$ with the corresponding energy density parameters defined at $a_{\rm eq}\equiv1$ simplifying to $\Omega_{\Lambda}=\Omega_{\rm m}=\nicefrac{1}{2}$.
Requiring $y(t_0)=\nicefrac{1}{2}$ and then normalizing results back to $a(t_0)\equiv1$ gives a present energy density parameter for the cosmological constant of
\begin{equation}
 \Omega_{\Lambda}=0.704 \,, \label{eq:omegalambda}
\end{equation}
which is consistent with current cosmological observations such as from the Planck satellite~\cite{Aghanim:2018eyx} and the Dark Energy Survey~\cite{Abbott:2017wau} at the $3\sigma$ and $1\sigma$ levels, respectively.
Note that the difference to the prediction of $\Omega_{\Lambda}=0.697$ in Ref.~\cite{Lombriser:2018aru}, found for an extension of the global sequestering mechanism, is due to the required collapse of the matter patches in that scenario at $t_{\rm max} = 4t_0$ whereas these structures are considered much longer-living here.

Finally, one observes that $y=\nicefrac{1}{2}$ is close to the epoch of minimal $y'$, which is another manifestation of a change from matter domination to comic acceleration and hence of the coincident energy densities.
This also implies, for instance, a current proximity to the maximal ratio of the density of critical mass shells to the critical background density~\cite{Dunner:2006rf}.
Cosmological coincidences have furthermore been pointed out at recombination, reionization, and the star-formation peak~\cite{Lombriser:2017cjy}.
It may be interesting to elaborate whether the framework presented here could establish a link between these observations or provide new approaches for other cosmological problems, leaving a range of further explorations to future work.

\psec{Conclusions}
%
The cosmological constant problem encompassing the weakly or non-gravitating vacuum energy, the late-time accelerated cosmic expansion, and the coincident current energy densities of the cosmological constant and matter remains a difficult puzzle to cosmology.
A new framework was presented here that proposes the additional variation of the Einstein-Hilbert action with respect to the quadratic Planck mass on top of the usual metric variation.
It offers the interpretation of identifying the Planck mass with a global Lagrange multiplier that imposes general relativistic dynamics on the metric prescribing the space-time for the matter fields.
The variation provides a constraint equation on the average Ricci scalar that acts to prevent vacuum energy from gravitating.
The evaluation of this constraint under consideration of the evolution of the inhomogeneous matter distribution in the Universe in form of disconnected matter cells representing ultimately isolated gravitationally bound structures yields a backreaction effect that self-consistently produces the cosmological constant of the background.
With the application of a uniform prior on the dimensionless physical size of these structures as a measure of likelihood determining our location in the cosmic history, one finds a mean expectation for the current energy density parameter of the cosmological constant of $\Omega_{\Lambda}=0.704$.
The result is in good agreement with current cosmological observations, giving rise to a late-time acceleration of the cosmic expansion and a coincident current energy density of matter.
Future analysis will reveal if the presented framework allows for a reinterpretation and possible unravelling of some cosmological obscurities such as coincidences identified for the epochs of recombination, reionization, and the star-formation peak with the times of equality between the energy densities of radiation, baryons, and the cosmological constant.
It may also motivate new approaches for other unresolved problems of cosmology.

%
This work was supported by a Swiss National Science Foundation Professorship grant (No.~170547).
Please contact the author for access to research materials.

\bibliographystyle{arxiv_physrev_mod}
\bibliography{ccproblem}

\def\eprinttmppp@#1arXiv:@{#1}
\providecommand{\arxivlink[1]}{\href{http://arxiv.org/abs/#1}{arXiv:#1}}
\def\eprinttmp@#1arXiv:#2 [#3]#4@{\ifthenelse{\equal{#3}{x}}{\ifthenelse{
\equal{#1}{}}{\arxivlink{\eprinttmppp@#2@}}{\arxivlink{#1}}}{\arxivlink{#2}
  [#3]}}
\providecommand{\eprintlink}[1]{\eprinttmp@#1arXiv: [x]@}
\renewcommand{\eprint}[1]{\eprintlink{#1}}
\providecommand{\eprintmod}[1][XXXX.XXXX]{\eprintlink{#1}}
\providecommand{\adsurl}[1]{\href{#1}{ADS}}
\renewcommand{\bibinfo}[2]{\ifthenelse{\equal{#1}{isbn}}{\href{http://cosmologist.info/ISBN/#2}{#2}}{#2}}
\begin{thebibliography}{27}
\expandafter\ifx\csname natexlab\endcsname\relax\def\natexlab#1{#1}\fi
\expandafter\ifx\csname bibnamefont\endcsname\relax
  \def\bibnamefont#1{#1}\fi
\expandafter\ifx\csname bibfnamefont\endcsname\relax
  \def\bibfnamefont#1{#1}\fi
\expandafter\ifx\csname citenamefont\endcsname\relax
  \def\citenamefont#1{#1}\fi
\expandafter\ifx\csname url\endcsname\relax
  \def\url#1{\texttt{#1}}\fi
\expandafter\ifx\csname urlprefix\endcsname\relax\def\urlprefix{URL }\fi

\bibitem{Riess:1998cb}
Supernova Search Team, A.~G. Riess {\em et~al.},
\newblock Astron. J. {\bf 116}, 1009 (1998).

\bibitem{Perlmutter:1998np}
Supernova Cosmology Project, S.~Perlmutter {\em et~al.},
\newblock Astrophys. J. {\bf 517}, 565 (1999).

\bibitem{Weinberg:1988cp}
S.~Weinberg,
\newblock Rev. Mod. Phys. {\bf 61}, 1 (1989).

\bibitem{Martin:2012bt}
J.~Martin,
\newblock Comptes Rendus Physique {\bf 13}, 566 (2012).

\bibitem{Aghanim:2018eyx}
Planck, N.~Aghanim {\em et~al.},
\newblock \eprintmod[arXiv:1807.06209].

\bibitem{Abbott:2017wau}
DES, T.~M.~C. Abbott {\em et~al.},
\newblock Phys. Rev. {\bf D98}, 043526 (2018).

\bibitem{Lombriser:2016yzn}
L.~Lombriser and N.~A. Lima,
\newblock Phys. Lett. {\bf B765}, 382 (2017).

\bibitem{Monitor:2017mdv}
Virgo, Fermi-GBM, INTEGRAL, LIGO Scientific, B.~P. Abbott {\em et~al.},
\newblock Astrophys. J. {\bf 848}, L13 (2017).

\bibitem{Lombriser:2017cjy}
L.~Lombriser and V.~Smer-Barreto,
\newblock Phys. Rev. {\bf D96}, 123505 (2017).

\bibitem{Kaloper:2016yfa}
N.~Kaloper, A.~Padilla and D.~Stefanyszyn,
\newblock Phys. Rev. {\bf D94}, 025022 (2016).

\bibitem{BeltranJimenez:2017tkd}
J.~B. Jim{\'e}nez, L.~Heisenberg and T.~Koivisto,
\newblock Phys. Rev. {\bf D98}, 044048 (2018).

\bibitem{Kaloper:2013zca}
N.~Kaloper and A.~Padilla,
\newblock Phys. Rev. Lett. {\bf 112}, 091304 (2014).

\bibitem{Kaloper:2015jra}
N.~Kaloper, A.~Padilla, D.~Stefanyszyn and G.~Zahariade,
\newblock Phys. Rev. Lett. {\bf 116}, 051302 (2016).

\bibitem{Wang:2017oiy}
Q.~Wang, Z.~Zhu and W.~G. Unruh,
\newblock Phys. Rev. {\bf D95}, 103504 (2017).

\bibitem{Kaloper:2016jsd}
N.~Kaloper and A.~Padilla,
\newblock Phys. Rev. Lett. {\bf 118}, 061303 (2017).

\bibitem{Henneaux:1984ji}
M.~Henneaux and C.~Teitelboim,
\newblock Phys. Lett. {\bf 143B}, 415 (1984).

\bibitem{Unruh:1988in}
W.~G. Unruh,
\newblock Phys. Rev. {\bf D40}, 1048 (1989).

\bibitem{Barrow:2010xt}
J.~D. Barrow and D.~J. Shaw,
\newblock Phys. Rev. Lett. {\bf 106}, 101302 (2011).

\bibitem{Shaw:2010pq}
D.~J. Shaw and J.~D. Barrow,
\newblock Phys. Rev. {\bf D83}, 043518 (2011).

\bibitem{Aurilia:1980xj}
A.~Aurilia, H.~Nicolai and P.~K. Townsend,
\newblock Nucl. Phys. {\bf B176}, 509 (1980).

\bibitem{Hawking:1984hk}
S.~W. Hawking,
\newblock Phys. Lett. {\bf 134B}, 403 (1984).

\bibitem{Bousso:2000xa}
R.~Bousso and J.~Polchinski,
\newblock JHEP {\bf 06}, 006 (2000).

\bibitem{Lombriser:2018aru}
L.~Lombriser,
\newblock \eprintmod[arXiv:1805.05918].

\bibitem{Dunner:2006rf}
R.~D{\"u}nner, P.~A. Araya, A.~Meza and A.~Reisenegger,
\newblock Mon. Not. Roy. Astron. Soc. {\bf 366}, 803 (2006).

\bibitem{Zwane:2017xbg}
N.~Zwane, N.~Afshordi and R.~D. Sorkin,
\newblock Class. Quant. Grav. {\bf 35}, 194002 (2018).

\bibitem{Brandenberger:2018fdd}
R.~Brandenberger, L.~L. Graef, G.~Marozzi and G.~P. Vacca,
\newblock Phys. Rev. {\bf D98}, 103523 (2018).

\end{thebibliography}

\end{document}